\documentclass{bmcart}

\usepackage{amsthm,amsmath,amsfonts}
\usepackage{algorithm}
\usepackage{algpseudocode}
\usepackage{graphicx}
\usepackage{caption}    
\usepackage{subfig} 
\graphicspath{ {./graphicspath/} }

\usepackage[utf8]{inputenc} %unicode support
\startlocaldefs
\endlocaldefs

\begin{document}

%%% Start of article front matter
\begin{frontmatter}

\begin{fmbox}
\dochead{} %%%% Audio Processing


\title{Music Recommendation Based on Audio Fingerprint}

%%%%%%%%%%%%%%%%%%%%%%%%%%%%%%%%%%%%%%%%%%%%%%
%%                                          %%
%% Enter the authors here                   %%
%%                                          %%
%% Specify information, if available,       %%
%% in the form:                             %%
%%   <key>={<id1>,<id2>}                    %%
%%   <key>=                                 %%
%% Comment or delete the keys which are     %%
%% not used. Repeat \author command as much %%
%% as required.                             %%
%%                                          %%
%%%%%%%%%%%%%%%%%%%%%%%%%%%%%%%%%%%%%%%%%%%%%%

\author[
  addressref={aff1},                   % id's of addresses, e.g. {aff1,aff2}
  corref={aff1},                       % id of corresponding address, if any
% noteref={n1},                        % id's of article notes, if any
  email={diego.ulloa13@hotmail.com}   % email address
]{\inits{D.S.U.}\fnm{Diego} \snm{Salda\~na-Ulloa}}
% \author[
%   addressref={aff1,aff2},
%   email={john.RS.Smith@cambridge.co.uk}
% ]{\inits{J.R.S.}\fnm{John R.S.} \snm{Smith}}

%%%%%%%%%%%%%%%%%%%%%%%%%%%%%%%%%%%%%%%%%%%%%%
%%                                          %%
%% Enter the authors' addresses here        %%
%%                                          %%
%% Repeat \address commands as much as      %%
%% required.                                %%
%%                                          %%
%%%%%%%%%%%%%%%%%%%%%%%%%%%%%%%%%%%%%%%%%%%%%%

\address[id=aff1]{%                           % unique id
  \orgdiv{},             % department, if any
  \orgname{},          % university, etc
  \city{},                              % city
  \cny{}                                    % country
}
% \address[id=aff2]{%
%   \orgdiv{Institute of Biology},
%   \orgname{National University of Sciences},
%   %\street{},
%   %\postcode{}
%   \city{Kiel},
%   \cny{Germany}
% }

%%%%%%%%%%%%%%%%%%%%%%%%%%%%%%%%%%%%%%%%%%%%%%
%%                                          %%
%% Enter short notes here                   %%
%%                                          %%
%% Short notes will be after addresses      %%
%% on first page.                           %%
%%                                          %%
%%%%%%%%%%%%%%%%%%%%%%%%%%%%%%%%%%%%%%%%%%%%%%

%%\note{Sample of title note}     % note to the article
%\note[id=n1]{Equal contributor} % note, connected to author

\end{fmbox}% comment this for two column layout

%%%%%%%%%%%%%%%%%%%%%%%%%%%%%%%%%%%%%%%%%%%%%%%
%%                                           %%
%% The Abstract begins here                  %%
%%                                           %%
%% Please refer to the Instructions for      %%
%% authors on https://www.biomedcentral.com/ %%
%% and include the section headings          %%
%% accordingly for your article type.        %%
%%                                           %%
%%%%%%%%%%%%%%%%%%%%%%%%%%%%%%%%%%%%%%%%%%%%%%%

\begin{abstractbox}

\begin{abstract} % abstract
 
This work combined different audio features to obtain a more robust fingerprint to be used in a music recommendation process. The combination of these methods resulted in a high-dimensional vector. To reduce the number of values, PCA was applied to the set of resulting fingerprints, selecting the number of principal components that corresponded to an explained variance of $95\%$. Finally, with these PCA-fingerprints, the similarity matrix of each fingerprint with the entire data set was calculated. The process was applied to 200 songs from a personal music library; the songs were tagged with the artists' corresponding genres. The recommendations (fingerprints of songs with the closest similarity) were rated successful if the recommended songs' genre matched the target songs' genre. With this procedure, it was possible to obtain an accuracy of $89\%$ (successful recommendations out of total recommendation requests).
% \parttitle{First part title} %if any
% Text for this section.

% \parttitle{Second part title} %if any
% Text for this section.
\end{abstract}

%%%%%%%%%%%%%%%%%%%%%%%%%%%%%%%%%%%%%%%%%%%%%%
%%                                          %%
%% The keywords begin here                  %%
%%                                          %%
%% Put each keyword in separate \kwd{}.     %%
%%                                          %%
%%%%%%%%%%%%%%%%%%%%%%%%%%%%%%%%%%%%%%%%%%%%%%


% MSC classifications codes, if any
%\begin{keyword}[class=AMS]
%\kwd[Primary ]{}
%\kwd{}
%\kwd[; secondary ]{}
%\end{keyword}

\end{abstractbox}
%
%\end{fmbox}% uncomment this for two column layout

\end{frontmatter}

%%%%%%%%%%%%%%%%%%%%%%%%%%%%%%%%%%%%%%%%%%%%%%%%
%%                                            %%
%% The Main Body begins here                  %%
%%                                            %%
%% Please refer to the instructions for       %%
%% authors on:                                %%
%% https://www.biomedcentral.com/getpublished %%
%% and include the section headings           %%
%% accordingly for your article type.         %%
%%                                            %%
%% See the Results and Discussion section     %%
%% for details on how to create sub-sections  %%
%%                                            %%
%% use \cite{...} to cite references          %%
%%  \cite{koon} and                           %%
%%  \cite{oreg,khar,zvai,xjon,schn,pond}      %%
%%                                            %%
%%%%%%%%%%%%%%%%%%%%%%%%%%%%%%%%%%%%%%%%%%%%%%%%

%%%%%%%%%%%%%%%%%%%%%%%%% start of article main body
% <put your article body there>

%%%%%%%%%%%%%%%%
%% Background %%
%%

\section*{Introduction}
The use of digital systems that operate with music tracks has increased considerably in recent years. Different tools exist for analyzing audio files, and applications focus on correctly identifying a musical record to provide additional services \cite{Cano}. Audio identification has become an essential part of different tasks on a day-to-day basis. This process is carried out employing audio fingerprints. In this scope, a fingerprint is defined as a unique representation of an audio track formed by different descriptive audio features. 

Audio fingerprinting techniques are sometimes confused with audio watermarking techniques. Audio watermarking aims to embed information in an audio track, for example, general information about the track for its correct identification. In this way, audio fingerprints can have many immediate applications, such as identification of copyrighted tracks, music recommendations, or voice recognition, to name a few \cite{book1}. An audio identification process through watermarking would involve directly extracting the embedded message. However, audio watermarking could introduce noise since the audio track information must be altered \cite{Cano}.

An audio identification system using fingerprints must have specific characteristics that make the system robust and capable of performing correct matches \cite{Cano}, among which we find: the number of accurate identifications, the security of the system against noisy signals, the granularity related to fragments audio files as short as possible, the independence of the audio formats, the scalability over the number of fingerprints in the system and the complexity related to the computational cost.

The audio identification system generally comprises an extraction block and a block that stores the information for later identification. A system like this can develop numerous applications; one of the most striking at present is the one related to music recommendation systems \cite{schedl}. The central idea relies on obtaining the fundamental characteristics of a musical piece (fingerprint) and recommending similar pieces.

In this way, the musical recommendation problem depends to a large extent on the construction of the fingerprint, that is, the set of techniques used to define the fingerprint of a song. The process generally involves extracting time-frequency features based on modifications to the Fourier transform \cite{bellitini}, and there are different methods to achieve this objective \cite{book2} \cite{haitsma} \cite{kim}. 

In this work, we combine different methods that extract information from an audio track based on its spectral, chromatic, and tempo features to extract the most significant amount of structural information and provide robustness to the process. Subsequently, we apply this procedure to a set of predefined songs to fully characterize them; the intention is to develop a recommendation system based on an individual's musical tastes.

The structure of this work is as follows: Section 1 deals with previous related works, section 2 introduces the theory for the extraction of spectral, chromatic, and tempo features given an audio signal; Section 3 presents the design proposal of the feature extraction system, and the recommendation process, Section 4 focuses on the experimental results taking into account the predefined data set; finally, the conclusions are detailed.

\section{Related Works}

The first version of audio fingerprinting was used in the 90s to detect advertisements in broadcast streams \cite{lourens}. This first version used the signal's energy information to match the signal waves' envelope. The objective focused on reusing transmissions and the automatic placement of other advertisements.

Currently, the processes to obtain the fingerprint of an audio signal focus on the extraction of structural features based on spectral analysis. For example, it is common to use variations of the Fourier transform, such as the Short Time Fourier Transform STFT \cite{bellitini}\cite{ramona}\cite{ramalingam}, which aims to extract the transformation coefficients associated with the frequencies of the signal. Unlike the standard Fourier transform, STFT operates on time windows, resulting in the final coefficients being associated with interval frequencies of the signal and not the complete signal. Another spectral technique used for fingerprinting is the Mel Frequency Cepstral Coefficients (MFCC) \cite{thiru}\cite{abu}\cite{khasana}. The MFCC is obtained by mapping the STFT coefficients to the Mel scale, defined as a perceptual scale based on pitches and frequencies perceptible by the human ear.

There are also features related to the pitch classes of Western music called chroma features. These variables capture the melodic characteristics of a piece of music. Different works use this type of variable \cite{kim}\cite{mioto}, and, like the MFCC, they work by calculating the STFT coefficients, combining them with binning techniques.

On the other hand, it is common to use features related to the tempo or beats of the audio signal. This feature type has been used as a complementary process in audio analysis to obtain fingerprints related to musical aspects such as tempo or rhythm \cite{ellis}\cite{kurth}. Generally, this approach works using the spectral parameters obtained by STFT to identify the onset of the signal corresponding to the beats' fundamental characteristics.

The types of features that we have mentioned correspond to spectral, chromatic, tempo, and Mel Scale-based. Some works have focused on combining some \cite{ramalingam}\cite{thiru} \cite{khasana} \cite{ellis} to obtain a more robust fingerprint capable of capturing more information about the audio signal and achieving correct identification. Despite the above, we did not find a work that is precisely dedicated to combining all the previous methods to obtain a fingerprint.

Regarding musical recommendation systems, some works focus exclusively on the similarities of songs based on spectral features \cite{badu}\cite{hanh}. Other works are in charge of combining this type of feature with more contextual information about the song, for example, the name, the duration, or the genre \cite{hanb}\cite{deldjoo}. The purpose is to create vector arrays that fully identify a piece of music, making it as distinctive as possible. With these arrays, songs can be compared to recommend similar items. Currently, music recommendation systems are used by large streaming music companies and use a combination of methods for constructing the features as well as different algorithms for comparing each element \cite{schedl}.

\section{Audio Features} 
Next, we will present the theory related to different types of audio features that will be used to develop a robust fingerprint. These features represent the audio signal's spectral, chromatic, Mel scale-based, and tempo characteristics. 

\subsection{Short Time Fourier Transform}

The Fourier transform is useful for analyzing different types of signals. This transformation aims to decompose the signal into frequency components $w \in \mathbb{R}$ associated with coefficients $d_w \in \mathbb{R}$. The coefficients indicate the degree to which a sinusoidal component with frequency $w$ adjusts the signal. In this way, the Fourier transform represents the signal in the frequency domain compared to its original form, which represents the time domain. For example, the Fourier transform of a musical extract indicates which notes (frequencies) have been played but not when they are played \cite{book2}.

In order to make use of the temporal information associated with a signal, the so-called short-time Fourier transform (STFT) is used. This transform type operates on small portions of the signal (time windows) compared to the original transform. Let $x \in \mathbb{R}$ be a discrete signal, $r:[0: N - 1] \rightarrow \mathbb{R}$ a time window function of length $N \in \mathbb{N}$, and a parameter $H \in \mathbb{N}$ called hop size that indicates the intervals in which the function $r$ moves along the signal; The STFT of the signal is given by

\begin{equation}
    \xi(m,k) = \sum^{N-1}_{n=0} x(n+mH)r(n)exp(-2\pi ikn/N),
\end{equation}

where $m \in \mathbb{Z}$ and $k\in[0:N/2]$. $\xi(m,k)$ is the $k^{\text{th}}$ fourier coefficient for the $m^{\text{th}}$ time frame. 

Since the results of STFT are complex numbers, we can transform those values into a real two-dimensional representation by squaring their magnitude

\begin{equation}
    \Theta (m,k) = |\xi(m,k)|^2.
\end{equation}

This is called the spectrogram and is helpful for visualizing the transformation. The horizontal axis represents the time, and the vertical axis represents frequency. 

\subsection{Mel Frequency Cepstral Coefficients}

This type of coefficient is based on the structural analysis of timbre. Initially, the coefficients were used for voice recognition tasks, and obtaining them involves mapping the frequencies to the Mel scale \cite{book2}. The Mel scale is a perceptual scale that tries to mimic the hearing tones of the human ear. It divides the frequencies into equidistant intervals on the final scale. The following function gives the frequency transformation to the Mel scale

\begin{equation}
    F_{\text{Mel}}(x) = 1127 \ \text{ln}(1+\frac{x}{700}).
\end{equation}

To compute the MFCC, a series of steps are used \cite{khasana} \cite{loganb} \cite{wibawa} that can be summarized as 1) calculate the STFT coefficients (as in the previous section); 2) make a mapping between the calculated coefficients and the Mel scale. A series of triangular filter banks along the Mel scale usually do this transformation. The objective is a binarization along the scale, depending on the number of filter banks used. The Mel coefficients have the following form

\begin{equation}
    S(n) = \sum^{N-1}_{k=0} |\xi(k)|^2 J_{m}(k),
\end{equation}

where $0\leq m \leq M-1$, $m$ is the number of triangular filters, and $J$ is the weight applied to the $k^{\text{th}}$ energy spectrum bin that contributes to the $m^{\text{th}}$  output band. This result is the energy spectrum of the Mel frequencies. 3) Finally, the Discrete Cosine Transform (DCT) is calculated on the previous coefficients to decorrelate them and obtain only a determined number of coefficients

\begin{equation}
    \Lambda_{m}(n) = \sum^{M-1}_{m=0} S(n) \ \text{cos}(\frac{\pi n(m-0.5)}{M}),
\end{equation}

where $n \in \mathbb{N}$ is the number of MFCC desired, in this work 13 coefficients were used because is a typical value according to \cite{khasana} \cite{loganb}.

\subsection{Chroma Coefficients}
This type of feature considers properties related to the harmony and melody of a signal. In order to obtain these features, tempered scales are considered, mainly the one related to Western music that divides the scale into 12 different tones $\{C,C\#,D,D\#,E,F,F\#,G,G\#,A,A\#,B\}$. To compute the chroma coefficients, we must calculate the STFT coefficients and obtain a log frequency spectrogram

\begin{equation}
    Y(n,p) = \sum_{k\in P(p)} |\xi(n,k)|^2,
\end{equation}

where $p\in [0:127]$ is the MIDI note number and represents the note's pitch, and $P(p)$ is a function that maps $k$ to the MIDI numbers. In short, the MIDI note number encodes the musical pitches along different octaves. Once this spectrogram is obtained, the chroma coefficients or chromagram are calculated by adding all the pitch coefficients that belong to the same chroma

\begin{equation}
    \Psi (n,c) = \sum_{p\in [0:127]:p \text{mod}12=c} Y(n,p),
\end{equation}

where $c\in[0:11]$.

\subsection{Tempo and Beat Features}

The beat corresponds to the pulse in an audio signal and is an additional characteristic that can be appreciated when listening to a song. The tempo is the rate at which the beats occur, corresponding to a fundamental aspect of music. Like chromatic coefficients, these features are related to the melodic aspects of an audio signal.

In order to identify the beats of a piece of music, one of the approaches used is beat tracking by dynamic programming. The goal is to generate a sequence of beat times that correspond to perceived onsets in the audio signal and that have a regular rhythm \cite{ellis2}. In order to do this, an objective function is used that combines a part related to the locations of the beats and another function that penalizes the deviations of irregular intervals

\begin{equation}
    \Xi({t_i}) = \sum^{N}_{i=1}O(t_i)+\alpha \sum^{N}_{i=2} F(t_i-t_{i-1},\tau_p)
\end{equation}

where ${t_i}$ is the sequence of $N$ beats, $O(t)$ is an onset strength envelope of the audio signal, $\alpha$ is a parameter to measure the importance of the penalty function $F$, and $\tau_p$ is an ideal beat spacing.

To calculate the best value of $\Xi$, we take the recursive relation

\begin{equation}
    \Xi^* (t) = O(t) + \max_{\tau=0,...,t} \{ \alpha F(t-\tau,\tau_p)+\Xi^*(\tau) \}
\end{equation}

In other words, the best value $\Xi$ for time t is the sum of the local onset strength plus the best value of $\Xi$ at time $\tau$. Since this function always operates with the preceding $\tau$ value (to get the instants of time $t$ that mark the beats), the iteration of this procedure is from the end towards the start of the signal.

%-----------------------------------------------------------
\begin{table}[h!]
\caption{Set of Songs}
  \begin{tabular}{cc}
    \hline
  Genre & Number of songs \\ \hline
  \\
   Metal Variants  & 28  \\
   Rock, Alternative, Indie, New Wave & 27  \\ 
   Ballad, Pop & 15  \\
   House, Electronic, Dance, Trance & 14  \\
   Synthpop, Electropop, Technopop & 12 \\
   Pop, Ballad, Rock & 12  \\
   Pop, Rock, Folk, Indie & 11  \\
   Jazz, Blues, Ballad & 10 \\
   Pop, Punk, Rock & 10  \\ 
   Electropop, Electronic, Hip hop & 9  \\
   Pop, Dance, Electropop & 9  \\
   Nortech, Electronic & 8  \\
   Rock, Pop, Urbano & 7  \\ 
   Classical, Soundtrack & 6  \\
   Pop, Indie, Rock & 6  \\ 
   Rock, Metal & 5  \\ 
   Pop, Rap, Dance & 4  \\
   Rock, Blues, Jazz & 4  \\
   Hip Hop, Rap, Rock & 3  \\   
   \\
      \hline 
  \end{tabular}
\label{table:setsongs}
\end{table}
%----------------------------------------------------------

\section{Audio Features and Music Recomendation}

In practical terms, the methods described in the previous section produce matrix coefficients that describe essential aspects of a musical composition. For example, the coefficients resulting from STFT, also called a spectrogram, give an overview of an audio signal since we obtain information about the frequencies that compose it. MFCCs provide information about timbre related to speech. Through the chroma coefficients, we can obtain properties of the harmony and melody related to the Western musical scale (chromagram). Finally, we can identify the exact moments corresponding to a song's beats (pulses) related to its tempo.

Although most of the mentioned methods start initially from the STFT results, each differs in the sequence of operations to obtain more information regarding an audio signal or song. This work focuses on obtaining the previously mentioned coefficients of a set of songs and combining them to obtain a more robust fingerprint that considers particular and essential aspects of a song.

To carry out this objective, we define the vector $\Omega$ as the vector that includes the coefficients of each of the methods described above, such that,

\begin{equation}
    \Omega = \bar{\Theta} \oplus \bar{\Lambda} \oplus \bar{\Psi} \oplus \bar{\Xi},
\end{equation}

 where $\bar{\Theta}$, $\bar{\Lambda}$, $\bar{\Psi}$ denotes the matrix of the mean values of the rows, $\bar{\Xi}$ is a vector of size of $\Omega$ with tempo values, and $\oplus$ indicates the concatenation of these matrices. With the mean values of the rows of each matrix, we reduce the dimension encapsulating the variations of each frequency along all the time. 
 
 Defining this vector or fingerprint focuses on condensing as much information as possible from an audio signal. However, due to the concatenation of the matrices, the resulting vector can have high dimensionality. Since the objective is to obtain musical recommendations, reducing the dimensionality of the vector is essential to reduce the complexity of the process.

One widely used technique to reduce a data set's dimensionality is principal component analysis (PCA) \cite{ringner}. This method creates linear combinations of the original variables called principal components. In this way, the principal components retain part of the original information (variance) of the original data. This reduction method aims to choose the number of principal components that retain a considerable percentage of the total variance of the original data \cite{ringner}.

Let $\Omega = \{\Omega^{\{1\}}, \Omega^{\{2\}}, ..., \Omega^{\{n\}}\}$ be the matrix formed by the union of different musical fingerprints and $\Omega'$ the resulting matrix of the principal components of $\Omega$, where $\Omega'$=\{$\Omega'_1, \Omega'_2, ..., \Omega'_n$\} correspond to the principal components of each of the musical fingerprints; we can compute the distance between fingerprints $i,j$ by some distance metric, like euclidean distance

\begin{equation}
    S_e =||\Omega_i'-\Omega_j'||
\end{equation}

For this case, the $S_e$ matrix corresponds to the distance matrix between each fingerprint. This matrix expresses the similarities between music tracks and is widely used in music recommendation systems \cite{schedl}. A music recommendation system aims to find the top K of the closest fingerprints (smallest distance) corresponding to the musical recommendations given a set of target songs. The target songs correspond to songs presumably added to a list of favorite songs. This type of recommendation is called content-based and is used to give the most similar items the user has liked \cite{book4}.

\begin{figure}[h]
\includegraphics[scale=0.45]{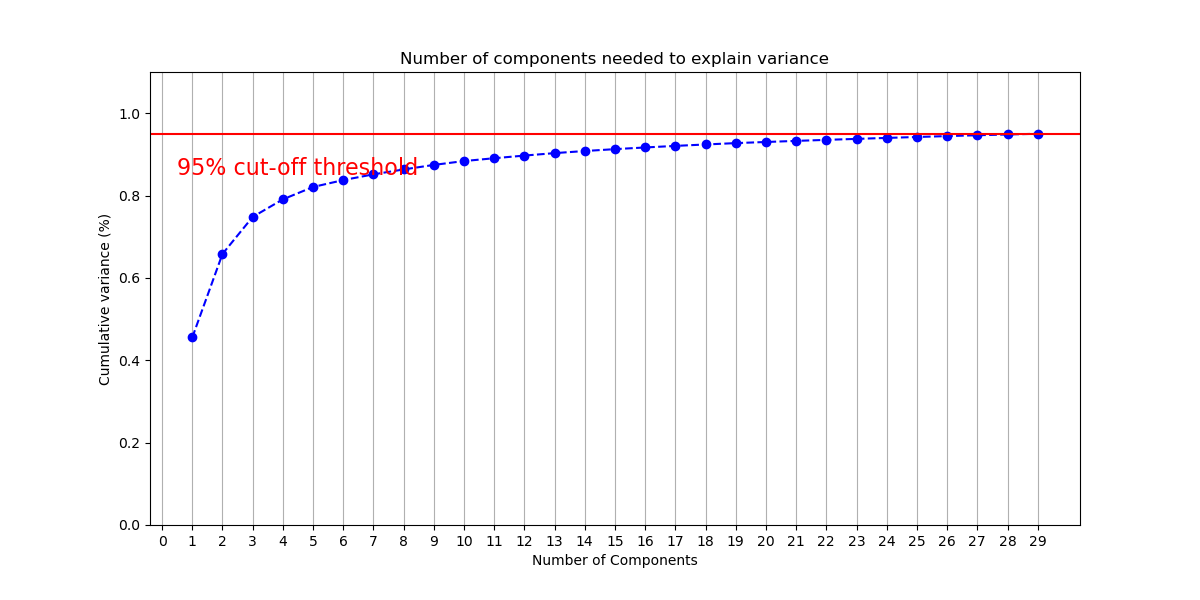}
\caption{The number of components needed to explain the $95\%$ of the variance}
\label{fig:variance}
\end{figure}

\section{Experimental Results}

For the experimental part, 200 songs from a personal music library composed of different genres and artists were used. Each song was tagged with the genres corresponding to the artist, obtained from Wikipedia. Table \ref{table:setsongs} shows the number of songs corresponding to a set of genres. Additionally, the bitrate of each audio file was kept constant at 128 kbps (CD quality). In order to compute the coefficient described in the previous sections, 60 seconds of audio were used starting at second 60, i.e., from the second 60 to the 120. Each audio file was processed in a mono signal. For the methods that use STFT, the sampling rate was set to 22050 Hz (samples per second), a window size of 2048, and a cosine window (hann).

The first part of the process focused on fingerprinting the sound files, as described in the previous section. In our case, each resulting vector (fingerprint) comprised 1062 coefficients. PCA was applied to each vector to reduce its dimensionality, and the number of components selected corresponded to a retained variance of $95\%$ Figure \ref{fig:variance}. This reduced the size of the vector from 1062 to 29, that is, a reduction of $97\%$ on the number of dimensions.

\begin{figure}[h]
%\centering
\includegraphics[scale=0.27]{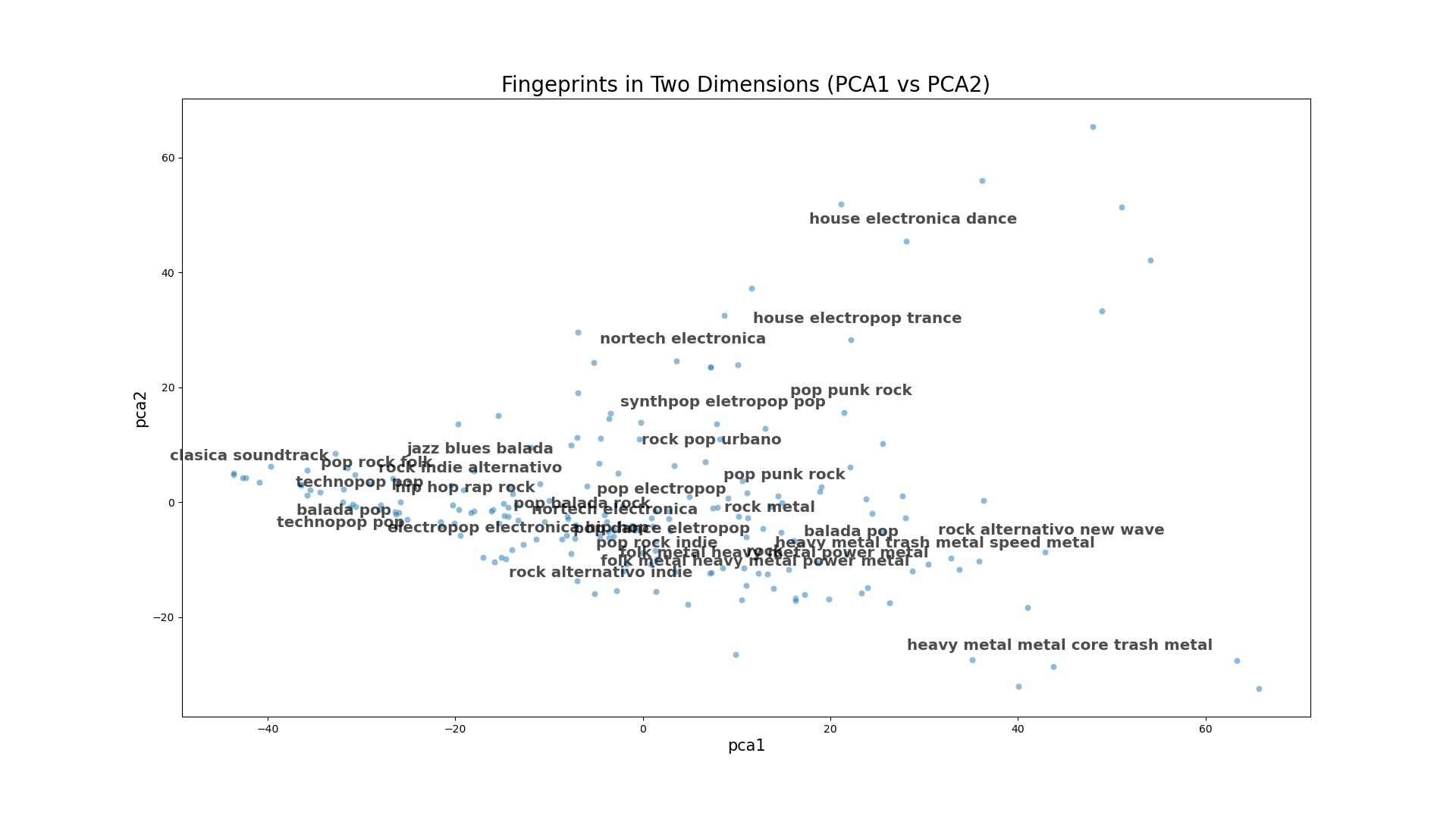}
\caption{The two principal components and the distribution of songs in this two dimensional space}
\label{fig:pc2}
\end{figure}

Figure \ref{fig:pc2} shows a graph of the two principal components of each vector song. It can be seen that the genres of the songs tend to cluster: in the center, we find the genres of rock, pop, alternative, and indie; in the lower right part, we find the metal variants; in the upper right part, the genres related to electronic music and in the central left part classical, ballads, jazz and indie. This shows us that chromatic, tempo, and spectral coefficients are useful in describing a music track.

The similarity between each pair of songs was computed using the 29 principal components for each song and the Euclidean distance, as described in the previous section. Then, the top-K of the most similar elements for each track was obtained using the similarity matrix. For this case, the top 3 most similar songs were retrieved, and a successful evaluation was set for any recommendation that corresponded to the genre of the target song. This type of evaluation system is commonly used within recommendation systems \cite{schedl}. With this consideration, it was possible to obtain an accuracy of $89\%$ (successful recommendations out of total recommendation requests), reaffirming the ability of the combination of chroma, tempo, MFCC, and STFT coefficients to obtain a robust fingerprint and to be able to categorize a song.

\section{Conclusions}

This work combined different audio features to obtain a robust musical fingerprint of a set of songs from a personal music library of 200 songs. Each song was tagged according to the musical genres of the artist. This fingerprint was later used in the process of recommending similar songs. The features used to define the fingerprint correspond to spectral, chromatic, tempo, and Mel scale-based features. Since each of the methods described above generally results in a matrix of values, the average values of the rows were obtained to capture the variation of each frequency throughout the signal. This process enabled obtaining a simplified vector (fingerprint) that combines the audio information described above. Because the size of the resulting vector contained 1062 elements, PCA was applied to the set of fingerprints of each song, and the 29 main components corresponding to a retained variance of $95\%$ were selected. With this information (PCA-fingerprints), the similarity matrix was calculated, defined as the matrix of values that reflect the similarity of each song with the rest of the data set. For each song, the top 3 most similar songs were obtained, and the recommendation was qualified as successful if the genre of any of the recommendations matched the genre of the target song. With this process, it was possible to obtain $89\%$ accuracy in the recommendation process, demonstrating that the union of different audio features allows for a more robust fingerprint that captures relevant song information.

\bibliographystyle{bmc-mathphys} % Style BST file (bmc-mathphys, vancouver, spbasic).
\bibliography{bmc_article}      % Bibliography file (usually '*.bib' )

\end{document}